\documentclass[aps,amsmath,showpacs,twocolumn]{revtex4}
\usepackage{graphicx}

\begin{document}

%\voffset=0.75truein

\title{Down-sizing Forever}

\author{Douglas Scott} \email{docslugtoast@phas.ubc.ca}
\author{Ali Frolop} \email{afrolop@phas.ubc.ca}
\affiliation{Department of Physics \& Astronomy\\
University of British Columbia,
Vancouver, BC, V6T 1Z1  Canada}

\date{1st April 2008}

\begin{abstract}
Evidence for cosmic down-sizing has been growing over the last decade.  It
is now clear that the major star-forming epoch for the largest galaxies
occurred {\it earlier\/} than for smaller galaxies.  This not only runs
counter to the popular hierarchical clustering picture, but points to an
even more radical revision of our ideas of the evolution of cosmological
structure.  Galaxies do not form at all.
\end{abstract}


\maketitle

\date{today}

\noindent

Mankind has long pondered the question of where everything came from,
and modern cosmology has been providing part of the answer.
It has become part of the conventional cosmological wisdom that the Universe
started off in
a very uniform state and that structure grew `hierarchically', in other
words smaller objects form earlier and aggregate into larger
objects at later times.  There would seem to be plenty of evidence for this
picture, e.g.~the smoothness of the Cosmic Microwave Background, 
the low overdensities in superclusters compared with clusters,
the evolution of the cluster mass function,
the power spectrum of galaxy clustering,
the distribution of Ly$\,\alpha$ clouds,
probes of dark matter potentials through cosmic shear,
direct observations of merging galaxies and the
apparent success of N-body simulations of Cold Dark Matter in explaining
all of these phenomena.

However, it is well-known that genuinely
unavoidable truths require only {\it one\/}
convincing argument to prove them.  Hence the fact that cosmologists rely
on so many pieces of evidence to support the hierarchical picture should
lead one to be quite skeptical.

So we should ask: are there any chinks in the armour of this cosmic hierarchy?
Indeed there are several, but we need only focus on one of them here.
The most revealing fact is that
the largest galaxies are full of very old stars, so that star-formation
appears to have progressed from larger galaxies to smaller ones.
The term chosen to describe this phenomenon is `down-sizing'
\cite{cowie}, in analogy
with the corporate metaphors of inflation, galaxy mergers and hostile
take-overs \cite{corporate}.
It now seems abundantly clear that the present time
in the history of the Universe is the domain of relatively small star-forming
galaxies, while the epoch of the giant galaxies was much earlier.  Indeed
a huge number of papers have studied various forms of down-sizing in the
last decade \cite{cattaneo}, and the evidence is now overwhelming.

Although it has become common to discuss how these findings are apparently
at odds with the basic premise of the hierarchical structure formation
paradigm, {\it no one\/} has had the foresight to take
these findings to their logical conclusion.

For decades researchers, struggling to understand how galaxies formed,
have tried to distinguish between 2 basic paradigms.  The first idea is where
galaxies collapse from one single immense cloud -- this being usually known as
the Monolithic model \cite{2001}.  The alternative is where galaxies
agglomerate from smaller sub-units, much like in the preparation of a
multi-layered snack food, and is usually called the ELS model \cite{ELS}.
This debate has entirely missed the point, just as on larger scales
cosmologists of the 1970s and 80s wasted their time arguing about `bottom-up'
versus `top-down' structure formation scenarios \cite{bottom}.

Modern astrophysicists have also been distracted by playing with so-called
`semi-analytic' \cite{anal} models for galaxy evolution.  By using scaling
relationships based on the observed properties of galaxies, one can
find astonishing agreement between the models and the observed properties of
galaxies.

However, it is now clear that astrophysicists studying the evolution of
galaxies have been focussing on entirely the wrong questions.
A radical rethinking of our cosmological ideas is required.
The reason why it has been hard to understand details of galaxy formation
is that galaxies {\it do not form at all!}  The lesson we should be taking
from the preponderance of evidence for down-sizing is that {\it galaxies
have actually been disappearing} for billions of years.

Anyone who can remember back to high school mathematics homework is aware that
the easiest of all mistakes to make is that of the slipped minus sign.
Physical cosmologists have been making the world's biggest minus sign
error in thinking that structure builds up over time, when in fact the very
opposite happens.  In keeping with the commercial metaphors we can think of
this not so much as `evaporation' or `vanishing', but more as `stream-lining'
or `rationalization' of galaxies \cite{layoff}

How could extragalactic researchers have made such a cosmic gaffe?  It is
well known that even Einstein had a `greatest blunder' in not appreciating
the importance of his own cosmological constant.  Moreover, no theorist
was smart enough to predict the expanding Universe, even although any modern
cosmologist would have stated that it was obvious if they had been around in 
1917 \cite{1917}.

But there were already clues missed from the early days when astronomers first
went looking for evidence of missing matter.
Despite evidence for dark matter from Zwicky as early
as 1933 \cite{zwicky}, some skeptics pointed out that galaxy clusters might
not be stable \cite{ambartzumian} -- in other words, although it was easy to
measure a velocity dispersion, it was fiendishly difficult to figure out
whether the galaxies were orbitting each other or flying apart (as now seems
to be the case).

So in our new extended down-sizing picture, the early history of
the Universe contained even bigger galaxies than exist today.
We can easily see this in the Supergalactic Plane, which
represents the last vestige of the Supergalaxy which once existed in
our own neighbourhood.  Even the name makes this obvious.

One might wonder about the even earlier Universe, and how to reconcile
super-down-sizing with the relative smoothness of the microwave sky?
The answer is clear.  It is often said that the last scattering surface
is like the cosmic photosphere, and hence looking at the microwave sky
is like looking at a star, except inside-out \cite{insideout}.  However,
this also misses the point -- the early Universe was in fact a giant
{\it galaxy}, not a star at all!

The origin of all structure then was this single Primordial Galaxy,
similar to the
`Ylem' proposed by George Gamow and collaborators in the 1940s.  The reason
that this fragmented and dissolved was probably akin to the `$-C$ field',
the negative of the continuous creation field proposed by Fred Hoyle and
collaborators in the 1960s.  How exactly the disintegration of galaxies
happens is not entirely clear.  We believe that there is probably a
stochastic element to it, as well as a gradual decrease in the sizes of
galaxies.  Whenever a particular galaxy's down-sizing stagnates for a while,
then there is likely to be an abrupt evaporation event -- we refer to this
as regulatory removal or a `reg-rem' event.  This explains why some galaxies
appear to be coalescing, when in fact it is the very opposite.

Another obvious question to ask is what happens in the far future?  Clearly the
evaporation of structures will continue forever, until we are left with
no galaxies, and presumably no stars, planets or people either.  The ultimate
fate of the down-sizing idea then is what might be call the `Big Fizzle'.

However, there may be ways out of this.  Theorists \cite{cite} have been keen
to find schemes for making cyclic models out of what seem otherwise
to be perfectly rational ideas.  So perhaps there is a way to reverse the
down-sizing and make the whole process repeat.  But a cyclic model need not go
on forever -- so perhaps we live in such a model, but there is only one
cycle \cite{unicycle}.  Or perhaps a combination with other, equally
plausible ideas \cite{us}
would help avoid the ultimate state of cosmic ennui that we predict.

Although our conclusions may be dramatic, perhaps we ourselves have also
been making blunders of astronomical proportions.
If, as we suggest, physical cosmologists have been making such an important
sign error, then maybe similar mistakes have been made elsewhere?
Perhaps overdensities are
really underdensities?  Perhaps dark matter is really bright?  Perhaps string
theory is really testable after all?
Perhaps the arrow of time points backwards?  Perhaps the Big Bang wasn't
really an explosion?  Perhaps people take their own ideas too seriously?
\cite{further}

%%%%%%%%%%%%%%%%%%%%%%%%%%%%%%%%%%%%%%%%%%%%%%%%%%%%%%%%%%%%%%%%%
%%%
%%%                     BIBLIOGRAPHY
%%%
%%%%%%%%%%%%%%%%%%%%%%%%%%%%%%%%%%%%%%%%%%%%%%%%%%%%%%%%%%%%%%%%%

\smallskip

%\newpage
%\vskip .75 in


\begin{thebibliography}{9}
\bibitem{corporate}
Similarly, we describe: black hole formation as `restructuring';
the LMC as being a `subsidiary' of the Milky Way Galaxy;
M31 and the Milky Way as forming a `limited partnership';
and elliptical galaxies as `not having much shareholder equity'.
\bibitem{bottom} Or their alternatives, `top-up', in which galaxies get a
little extra mass late in their evolution, or `bottom-down' in which galaxies
just sit there and do nothing.
\bibitem{anal} Some wiseacres have pointed out that semi-`analytic' is
`ytic', while some others hold with a more retentive opinion of this
endeavour.
\bibitem{layoff} And the loss in matter which galaxies experience over time
is called a `mass lay-off'.
\bibitem{1917} And new what we now know.
\bibitem{cite} Presumably because just one period of eternity might not be
enough for their work to gain the fame it deserves.
\bibitem{further}
And perhaps the third paper in a series is just overdoing it?
Further such suggestions should be sent to the second author, Dr.~Frolop.
\end{thebibliography}
\end{document}